\documentclass[aip,
 mph,%
 amsmath,
 amssymb,
 superscriptaddress,
 floatfix,
 preprint,
 eqsecnum,%
 numerical,%
]{revtex4-1}

\usepackage{graphicx}
\DeclareGraphicsExtensions{.eps,.pdf}
\usepackage[caption=false
]{subfig}
\usepackage{lineno}
\usepackage{upgreek}

\usepackage{epsfig}
\usepackage{epstopdf}
\usepackage{lipsum}
\usepackage{wrapfig}
\usepackage{wasysym}
\usepackage{threeparttable,booktabs,siunitx}
\usepackage{booktabs}
\usepackage{lmodern}
\usepackage{longtable}
\usepackage{multirow}
\usepackage{amsmath}
\usepackage[OT2,T1]{fontenc}
\usepackage[bookmarks=false]{hyperref}
\usepackage{color}
\hypersetup{pdfauthor={},pdfborder=0 0 0,colorlinks=true,citecolor=blue,linkcolor=blue,urlcolor=blue,}

\DeclareSymbolFont{cyrletters}{OT2}{wncyr}{m}{n}
\DeclareMathSymbol{\comb}{\mathalpha}{cyrletters}{"58}

\newcommand{\ben}{\begin{eqnarray}\displaystyle}
\newcommand{\een}{\end{eqnarray}}

\begin{document}

\title{Model-driven CT reconstruction algorithm for nano-resolution X-ray phase contrast imaging}

\author{Yuhang Tan}
 \thanks{Yuhang Tan and Xuebao Cai have made equal contributions to this work and are considered as the first authors.}
\affiliation{Research Center for Medical Artificial Intelligence, Shenzhen Institute of Advanced Technology, Chinese Academy of Sciences, Shenzhen, 518055, China}
\author{Xuebao Cai}
 \thanks{Yuhang Tan and Xuebao Cai have made equal contributions to this work and are considered as the first authors.}
\affiliation{Institute of Electronic Paper Displays, South China Academy of Advanced Optoelectronics, South China Normal University Guangzhou 510006, China}
\author{Ting Su}
\affiliation{Research Center for Medical Artificial Intelligence, Shenzhen Institute of Advanced Technology, Chinese Academy of Sciences, Shenzhen, 518055, China}%
\author{Ryosuke Ueda}
\affiliation{Institute of Multidisciplinary Research for Advanced Materials, Tohoku University, 2-1-1 Katahira, Aoba-ku, Sendai, 980-8577, Japan}%
\author{Dong Liang}
\affiliation{Research Center for Medical Artificial Intelligence, Shenzhen Institute of Advanced Technology, Chinese Academy of Sciences, Shenzhen, 518055, China}%
\affiliation{Paul C Lauterbur Research Center for Biomedical Imaging, Shenzhen Institute of Advanced Technology, Chinese Academy of Sciences, Shenzhen 518055, China}
\author{Hairong Zheng}
\affiliation{Paul C Lauterbur Research Center for Biomedical Imaging, Shenzhen Institute of Advanced Technology, Chinese Academy of Sciences, Shenzhen 518055, China}
\author{Peiping Zhu}
\affiliation{Institute of High Energy Physics, Chinese Academy of Sciences, Beijing 100049, China}%
\author{Atsushi Momose}%
 \thanks{Authors to whom correspondence should be addressed: Atsushi Momose (atsushi.momose.c2@tohoku .ac.jp), Yongshuai Ge (ys.ge@siat.ac.cn).}
 \affiliation{Institute of Multidisciplinary Research for Advanced Materials, Tohoku University, 2-1-1 Katahira, Aoba-ku, Sendai, 980-8577, Japan}
\author{Yongshuai Ge}%
 \thanks{Authors to whom correspondence should be addressed: Atsushi Momose (atsushi.momose.c2@tohoku .ac.jp), Yongshuai Ge (ys.ge@siat.ac.cn).}
 \affiliation{Research Center for Medical Artificial Intelligence, Shenzhen Institute of Advanced Technology, Chinese Academy of Sciences, Shenzhen, 518055, China}
\affiliation{Paul C Lauterbur Research Center for Biomedical Imaging, Shenzhen Institute of Advanced Technology, Chinese Academy of Sciences, Shenzhen 518055, China}%

\date{\today}

\begin{abstract}
The low-density imaging performance of a zone plate based nano-resolution hard X-ray computed tomography (CT) system can be significantly improved by incorporating a grating-based Lau interferometer. Due to the diffraction, however, the acquired nano-resolution phase signal may suffer splitting problem, which impedes the direct reconstruction of phase contrast CT (nPCT) images. To overcome, a new model-driven nPCT image reconstruction algorithm is developed in this study. In it, the diffraction procedure is mathematically modeled into a matrix $\mathbf{B}$, from which the projections without signal splitting can be generated invertedly. Furthermore, a penalized weighed least-square model with total variation (PWLS-TV) is employed to denoise these projections, from which nPCT images with high accuracy are directly reconstructed. Numerical and physical experiments demonstrate that this new algorithm is able to work with phase projections having any splitting distances. Results also reveal that nPCT images with higher signal-to-noise-ratio (SNR) would be reconstructed from projections with larger signal splittings. In conclusion, a novel model-driven nPCT image reconstruction algorithm with high accuracy and robustness is verified for the Lau interferometer based hard X-ray nano-resolution phase contrast imaging.
\end{abstract}

\maketitle

\section{Introduction}
X-ray CT imaging with nano-resolution is a powerful technique in detecting the three-dimensional (3D) ultra-fine structures inside the object. For low-density objects, however, conventional X-ray imaging may not be able to provide sufficient image contrast due to the subtle variations of the absorption coefficients between two different materials. To overcome such difficulty, the grating based X-ray Lau interferometer was coupled with the zone plate hard X-ray CT imaging system~\cite{momose_2009}. By doing so, the phase information that carries better image contrast can be retrieved. When a Lau interferometer is incorporated into a nPCT imaging system~\cite{momose_2009}, however, both theory and experiments have shown that strong signal splitting would appear due to the dramatically magnified X-ray beam diffraction by the zone plate~\cite{momose_2009,Yang_2022}. Clearly, such signal splitting would hinder the direct reconstructions of nPCT images~\cite{takano2019improvement} from the acquired projections.

\begin{figure*}[t]
  \centering
  \includegraphics[width=0.85\linewidth]{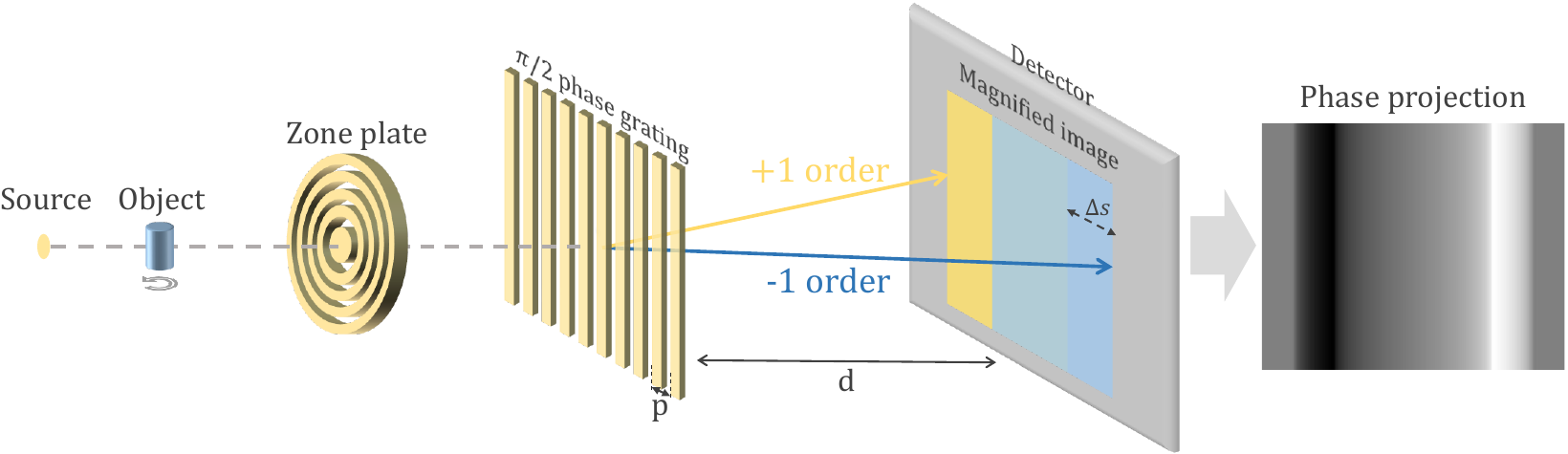}
  \caption{Illustration of a hard X-ray nano-resolution computed tomography (CT) system integrated with a zone plate and a Lau interferometer. The X-ray beam diffracts and splits (+1 and -1 orders) on the $\uppi$/2 phase grating, resulting in signal overlapping for the phase contrast projection.}
  \label{fig:imaging_process}
  \end{figure*}

To overcome, Takano \textit{et al.} proposed an image deconvolution based nPCT reconstruction algorithm~\cite{takano2019improvement}. In it, the phase signal recovery is achieved through a combination of signal post-processing procedures such as deconvolution and Fourier transformation. By doing so, projections without signal splitting can be obtained to allow the following reconstructions of nPCT images. To mitigate the phase wrapping artifacts, Ueda \textit{et al.} recently proposed a three-wave interference model to iteratively reconstruct the nPCT images from the measured beam intensities~\cite{ueda2021reconstruction}. In this study, a model-driven nPCT image reconstruction algorithm is proposed. Similarly as has been done by Ueda \textit{et al.}, the beam diffraction is mathematically modeled. In particular, it is expressed by an operator $\mathbf{B}$, from which the projections without signal splitting can be recovered inversely. Additionally, a penalized weighed least-square model~\cite{fessler} with total variation (PWLS-TV) is employed to denoise these projections. Finally, nPCT images with improved signal-to-noise ratio (SNR) are reconstructed via the filtered-backprojection (FBP) algorithm.

\section{Signal model}
In the following discussions, it is assumed that the nPCT imaging system comprises a X-ray source (together with a condenser), an object, a zone plate, a $\uppi$/2 phase grating, and a detector, see the illustration in Fig.~\ref{fig:imaging_process}. Note that the zone plate is used to magnify the image of the object, and the $\uppi$/2 phase grating is used to encode the phase information of the X-ray beam. Assuming the complex refractive index of the object is denoted as $n(x)=1-\delta(x)+i\beta(x)$, thus, the splitted phase signal $\Phi$ is expressed as~\cite{momose_2009,Yang_2022}:
\begin{equation}
  \begin{aligned}
    &\Phi (x) =  \varphi(x+\frac{\Delta s}{2})-\varphi(x-\frac{\Delta s}{2}),\\  
\end{aligned}
\label{eq:dpc_formula}
\end{equation}
where
\begin{equation}
  \begin{aligned}
    &\varphi(x) = k \int \delta(x) dz\\
\end{aligned}
\label{eq:absorption_dpc_par}
\end{equation}
denotes the phase signal before splitting, $\Delta s= \frac{2 \lambda d}{p}$ denotes the diffraction induced splitting distance, $k$ and $\lambda$ represent the wave number and wavelength, respectively, $d$ denotes the distance between phase grating and detector, and $p$ is the period of phase grating. Be aware that the phase signal $\Phi$ is an opposite-superposition (-) of the paired diffraction signals $\varphi$ in Eq.~(\ref{eq:dpc_formula}). Due to the non-zero splitting $\Delta s$, as a consequence, the phase signal $\Phi$ can not be directly utilized to reconstruct nPCT images.

\section{Reconstruction Algorithm}
\begin{figure}[t]
  \centering
  \includegraphics[width=1.0\linewidth]{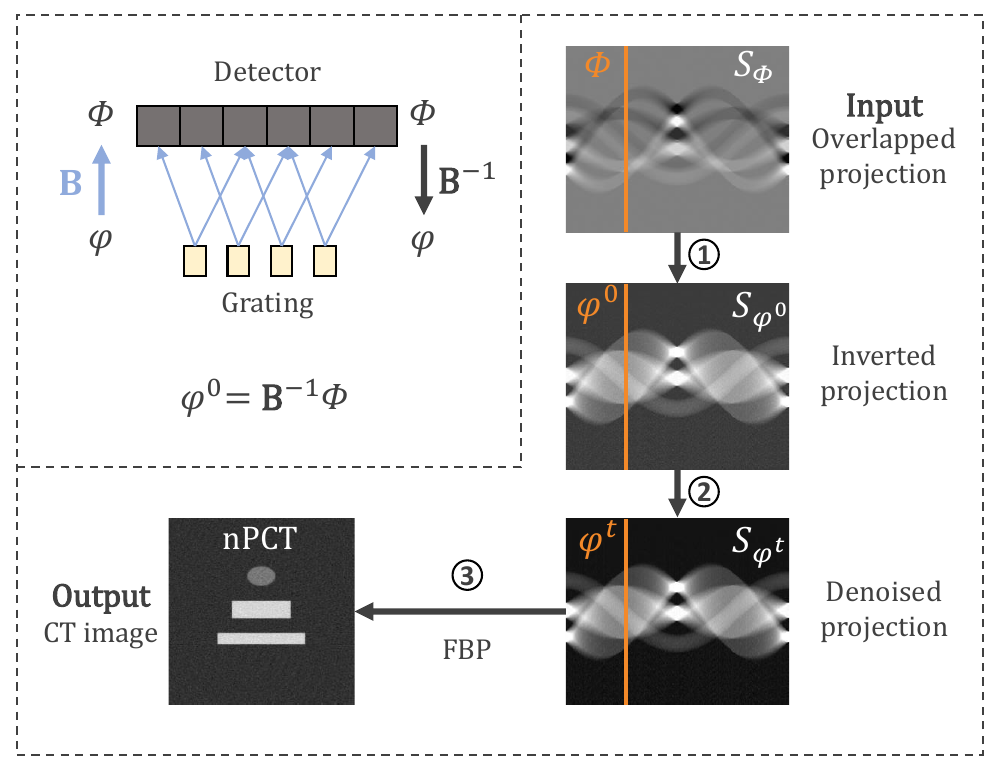}
  \caption{Workflow of the proposed nPCT image reconstruction algorithm. In step one, the overlapped projection $S_{\Phi}$ is converted to the unsplitted projection $S_{\varphi^{0}}$ using $\mathbf{B}^{-1}$. In step two, image $S_{\varphi^{0}}$ is iteratively denoised according to Eq.~(\ref{eq:gradient_descent}) for all projection views. In the final step, nPCT images are reconstructed from $S_{\varphi^{t}}$ using the FBP algorithm. The highlighted line denotes the phase projection acquired at a certain view.}
  \label{fig:overlap_rec_pro}
  \end{figure}

Mathematically, the signal diffraction, i.e., signal splitting, can be expressed by a matrix $\mathbf{B}$. With it, the phase projection $\Phi$ in Eq.~(\ref{eq:dpc_formula}) can be rewritten as:
\begin{equation}
  \Phi = {\rm \mathbf{B}} \varphi + n_{\varphi},
\label{eq:phi_matrix}
\end{equation}
where $n_{\varphi}$ represents the signal noise. According to Eq.(~\ref{eq:dpc_formula}), matrix $\mathbf{B}$ is defined as:
\begin{equation}
  \begin{aligned}
    & \mathbf{B}=\mathbf{b}_{m \times m} +  \boldsymbol{O}, \\
\label{eq:B_matrix}  
  \end{aligned}
\end{equation}
where
\begin{footnotesize}
\begin{center}
\begin{equation}
  \begin{aligned}
    & \mathbf{b}_{m \times m} = \left[\begin{array}{cccccccc}
      0 & \cdots & 0 & -1_{i,i+\Delta}^{i \leq \Delta} & 0 & \cdots & \cdots  & 0 \\
      \vdots & \vdots & \vdots & \vdots & \ddots & \vdots & \vdots  & \vdots \\
      0 & \cdots & 1_{i,i-\Delta}^{\Delta<i<=m-\Delta} & \cdots & \cdots & -1_{i,i+\Delta}^{\Delta<i<=m-\Delta}  & \cdots & 0 \\
      \vdots & \vdots & \vdots & \ddots & \vdots & \vdots & \vdots  & \vdots \\
      0 & \cdots & \cdots &  0 & 1_{i,i-\Delta}^{i>m-\Delta} & 0 & \cdots & 0
      \end{array}\right], 
\label{eq:Bdpc_matrix}  
  \end{aligned}
\end{equation}
\end{center}
\end{footnotesize}
matrix $\boldsymbol{O}=diag(\gamma,\gamma,\cdots,\gamma)$ ($\gamma=10^{-12}$) guarantees the invertibility of $\mathbf{B}$, index $m$ refers to the number of detector elements, $\Delta=\frac{\Delta s}{2w}$ denotes the splitting of half of signal separation in terms of the number of detector elements with width $w$, $i$ and $j$ represent the indices of row and column in the matrix, respectively. The work flow of $\mathbf{B}$ is illustrated in Fig.~\ref{fig:overlap_rec_pro}. If ignoring signal noise $n_{\varphi}$, essentially, the phase signal $\varphi$ in Eq.~(\ref{eq:phi_matrix}) without splitting can be solved by $\mathbf{B}^{-1}\Phi$, see Fig.~\ref{fig:overlap_rec_pro}. 

The analytical inversion of Eq.~(\ref{eq:phi_matrix}) would dramatically boost the image noise. To mitigate, the penalized weighted least-square model with total variation (PWLS-TV) is applied. PWLS-TV can effectively enhance the image quality, i.e., signal-to-noise-ratio (SNR), while preserving the edges and details~\cite{fessler}. In the case of nPCT imaging, the objective function is formulated as~\cite{fessler}:
\begin{equation}
{\varphi}^*=\underset{\varphi \geq 0}{\arg \min }\left\{\left(\Phi-\mathbf{B} \varphi\right)^{\mathbf{T}} \varLambda ^{-1}\left(\Phi-\mathbf{B} \varphi \right)+ \alpha \mathbf{R}_{\mathrm{TV}}(\varphi)\right\},
\label{eq:objective_function}
\end{equation}
where $\alpha$ denotes a smoothing parameter that controls the consistency between the estimation and measurement, $\varLambda $ denotes a diagonal matrix with the $q^{th}$ detector element $\sigma_{\varphi_{q}}^{2}$ equals to~\cite{chen2011scaling}:
\begin{equation}
  \sigma_{\varphi_{q}}^2=\frac{2}{\epsilon^2 \sum\limits_{k=1}^{M} {I}_{q}^{(k)}}.
  \label{eq:Sigma}
\end{equation}
Herein, $\epsilon$ represents the response efficiency of the interferometer and was set to 0.5 in this study, $M$ denotes the total number of phase steps, and ${I}_{q}$ represents the detected photons in the $q^{th}$ detector element.
In Eq.~(\ref{eq:objective_function}), the prior term $\mathbf{R}_{\mathrm{TV}}(\varphi)$ is defined as:
\begin{equation}
  \mathbf{R}_{\mathrm{TV}}(\varphi)=\sum_{q} \sqrt{\left(\varphi_{q}-\varphi_{q-1}\right)^2+\upsilon},
 \end{equation}
where $\upsilon$ is a small constant used for keeping $\mathbf{R}_{\mathrm{TV}}(\varphi)$ differentiable with respect to image intensity. 

The gradient descent method is used to iteratively solve the above optimization problem. Particularly,
\begin{equation}
  \small
  {\varphi}^{t}={\varphi}^{(t-1)}-\eta^{(t-1)}\left(\mathbf{B}^{\mathbf{T}}\left(\varLambda^{-1}\left(\mathbf{B} {\varphi}^{(t-1)}-\Phi\right)\right)\right)-\tau \frac{\nabla \mathbf{R}_{\mathrm{TV}}\left({\varphi}^{(t-1)}\right)}{\left\|\nabla \mathbf{R}_{\mathrm{TV}}\left({\varphi}^{(t-1)}\right)\right\|},
  \label{eq:gradient_descent}
\end{equation}
in which $\varphi^{t}$ denotes the optimized phase image obtained from the $t^{th}$ iteration, $\nabla\mathbf{R}_{\mathrm{TV}}\left({\varphi}^{(t-1)}\right)$ denotes the gradient of the regularizer, $\tau = 0.02$ denotes the update step size, and $\eta^{(t-1)}$ denotes the gradient step size~\cite{kawata1985constrained}:
\begin{equation}
  \eta^{t-1}=\frac{G^T G}{(\mathbf{B} G)^T\left(\varLambda^{-1}(\mathbf{B} G)\right)} \text { with } G \stackrel{\Delta}{=} \mathbf{B}^T\left(\varLambda^{-1}\left(\mathbf{B} \varphi^{t-1}-\Phi\right)\right).
\end{equation}
By repeating the signal processing procedure in Eq.~(\ref{eq:gradient_descent}) for all projection views, the unsplitted and denoised sinogram $S_{\varphi^{t}}$ would be obtained, see more details in Fig.~\ref{fig:overlap_rec_pro}.

\section{Numerical experiments and results}
\begin{figure}[h!]
  \centering
  \includegraphics[width=1.0\linewidth]{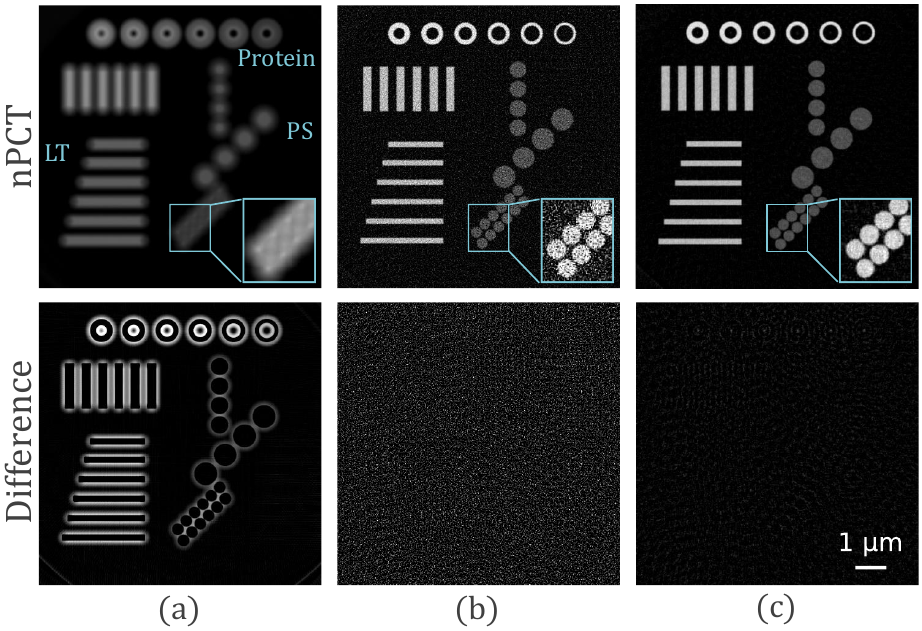}
  \caption{Comparison results with splitting distance $\Delta s$=200 nm. The nPCT images are reconstructed from (a) $S_{\Phi}$, (b) $S_{\varphi^{0}}$, (c) and $S_{\varphi^{t}}$. The display window is [0, 1.2 $\times$ $10^{-5}$]. The differences between the reconstructed nPCT images and the ground truth are shown below with a display window of [0, 4.5 $\times$ $10^{-6}$]. The scale bar denotes 1$\upmu$m.}
  \label{fig:fbp_overlap_alg}
  \end{figure}

Numerical experiments were conducted to verify the proposed image reconstruction algorithm. First of all, a two-dimensional (2D) phantom was simulated, see Fig.~\ref{fig:fbp_overlap_alg}. The numerical phantom has a size of 512$\times$512 with pixel dimension of 10 nm. It consists three groups of circles, rings and bars. They are made of three low-density materials: Lung tissue (LT), Protein and Polystyrene (PS). All the numerical simulations were performed on our previously developed X-ray nPCT imaging platform~\cite{Yang_2022}. In brief, the X-ray beam energy was 8.04 keV, 10000 X-ray photons were incident on the sample with Poisson noise, and the overall system magnification was 650. In total, 720 projections were captured by a 0.5 degree angular interval within a full rotation. The 1D detector array has a size of 1$\times$600, and each element has a dimension of 6.5 $\mu$m.

To compare, nPCT images were reconstructed from three different approaches. First, the acquired phase projections $S_{\Phi}$ containing signal splitting were reconstructed directly without any post-processing. Second, nPCT images were reconstructed from the analytically inverted phase signal $S_{\varphi^0}$ before image denoising. Third, nPCT images were reconstructed using the PWLS-TV denoised phase signal $S_{\varphi^t}$. Herein, the parallel CT imaging geometry was assumed, and the conventional FBP algorithm with Hilbert filter kernel for $S_{\Phi}$ and ramp filter kernel for $S_{\varphi^0}$ and $S_{\varphi^t}$ was implemented. 
Results are presented in Fig.~\ref{fig:fbp_overlap_alg}. As seen, the direct reconstruction would lead to very blurry nPCT images. As a contrary, the inversion of phase signal would significantly enhance the image sharpness. Moreover, the PWLS-TV approach can further reduce the image noise, see the highlighted region-of-interests (ROIs). The residual differences with respect to the numerical phantom are depicted in the bottom.

  \begin{figure}[h!]
    \centering
    \includegraphics[width=1.0\linewidth]{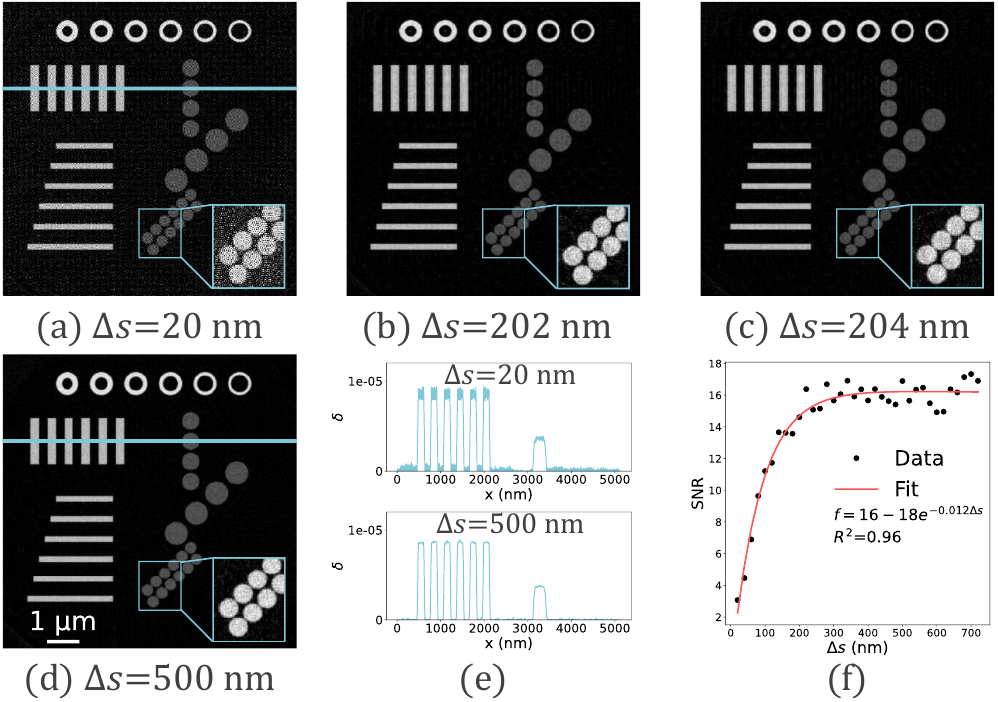}
    \caption{Results for different separation distance $\Delta s$ with pixel dimension of 10 nm. (a) $\Delta s$=20 nm, (b) $\Delta s$=202 nm, (c) $\Delta s$=204 nm, (d) $\Delta s$=500 nm. (e) The profile of highlighted line in (a) and (d). (f) Relationship between SNR and $\Delta s$ for nPCT. The scale bar denotes 1 $\upmu$m, and the gray scales of all images are identical from 0 to 1.2 $\times$ $10^{-5}$.}
    \label{fig:deltas_influence}
    \end{figure}
Additionally, the impact of the signal splitting $\Delta s$ to the quality of the reconstructed nPCT images was also investigated quantitatively. Images in Fig.~\ref{fig:deltas_influence} display the reconstructed nPCT results at four different splitting distances: $\Delta s$=20, 202, 204 and 500 nm. For $\Delta s$=202 nm and $\Delta s$=204 nm, which correspond to non-integer ($\Delta$=10.1 and $\Delta$=10.2) number of detector elements, the element $b^{}_{i,j}$ in matrix $\mathbf{b}_{m \times m}$ used in Eq.~(\ref{eq:B_matrix}) is defined as:
\begin{footnotesize}
  \begin{center}
  \begin{equation}
    \begin{aligned}
 b^{}_{i,j} = \begin{cases}
        f-1, & j=i+\Delta\\
        -f, & j=i+\Delta+1\\
        f, & j=i-\Delta-1\\
        1-f, & j=i-\Delta\\
        0, & \text{otherwise}
        \end{cases},
  \label{eq:Bdpc_Ba_matrix}  
    \end{aligned}
  \end{equation}
  \end{center}
  \end{footnotesize}
where $f$ corresponds to the fractional part of 
$\Delta$. For instance, if $\Delta$=10.1, then $f$=0.1. Moreover, the relationship between image SNR and $\Delta s$ is plotted as well, see Fig.~\ref{fig:deltas_influence}(f). Interestingly, it is found that larger signal splitting $\Delta s$ leads to nPCT images with improved SNR. As $\Delta s$ gets large enough, SNR tends to become constant. It is guessed that this might be related to the noise correlation induced by the signal overlapping. Currently, we do not have thorough explanations and further investigations are needed in the future.

\begin{figure}[t]
  \centering
  \includegraphics[width=1.0\linewidth]{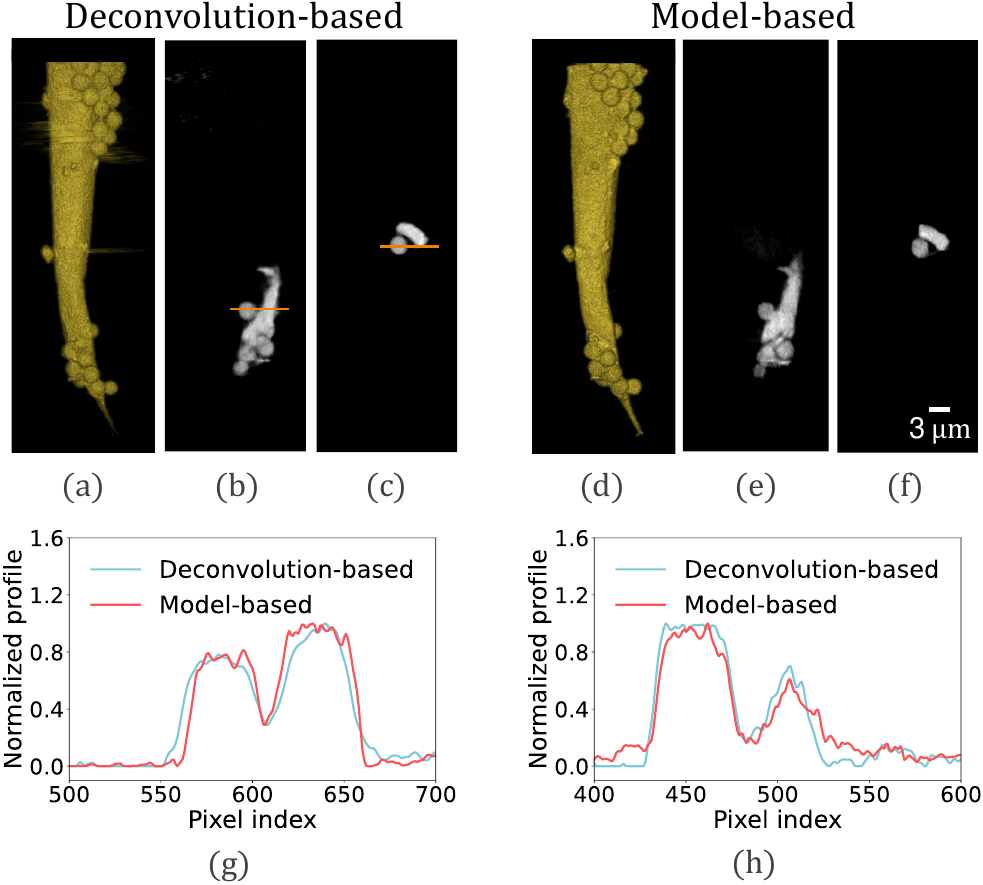}
  \caption{Experimental imaging results of the plastic fiber sample. (a)-(c) Images are obtained from the deconvolution based phase retrieval method, (d)-(f) Images are obtained from the newly developed model based phase retrieval method, (g) Line profile comparison results of the highlighted region in (b) and (e), (h) Line profile comparison results of the highlighted region in (c) and (f). The scale bar denotes 3 $\upmu$m.}
  \label{fig:experiment_result}
  \end{figure}
  
\section{Physical experiments and results}
Physical experiments were conducted on a dedicated nano-resolution hard X-ray phase contrast imaging system (ZEISS Xradia 800 Ultra systeincorporated with Lau interferometer~\cite{takano2019development}). The X-ray beam energy was 8.04 keV, the distance from the $\uppi$/2 phase grating (2.4 $\upmu$m period) to the detector was 559 mm. For this experimental system, it was found that $\Delta s=7.2~ \upmu$m. In total, 361 angular projections were captured with 0.5 degree interval. The phase stepping number was 6. The scanned object was a plastic fiber adhered with some 3.0 $\upmu$m PS spheres.



The reconstructed nPCT results are shown in Fig~\ref{fig:experiment_result}. Compared to the deconvolution based algorithm~\cite{takano2019improvement}, the newly proposed model based algorithm is able to generate almost identical nPCT images, see the 3D rendered volumes in Fig~\ref{fig:experiment_result}(a) and (d). The sagittal slices are presented in Fig~\ref{fig:experiment_result}(b) and (e), and the axial slices are presented in Fig~\ref{fig:experiment_result}(c) and (f). In addition, line profiles are also compared, see the plots in Fig~\ref{fig:experiment_result}(g) and (h). Results demonstrate the high fidelity of this newly proposed model based nPCT image reconstruction algorithm. Note that the geometric distortion calibrations used by the two algorithms are slightly mismatched, as a consequence, the nPCT images on the same slices are not identical in Fig~\ref{fig:experiment_result}.

\section{Discussions and Conclusion}
This study proposes a new model-driven nPCT image reconstruction algorithm to solve the signal splitting problem occurred in the nPCT imaging system coupled with a zone plate and a Lau interferometer. In it, the diffraction induced signal splitting is modeled by an operation matrix $\mathbf{B}$, which can be well defined and analytically inverted. Thus, the phase signal before splitting can be immediately recovered. Additionally, the PWLS-TV method is developed to further enhance the SNR of the recovered unsplitting phase signal. Eventually, nPCT images of high quality are reconstructed via the FBP method. Both numerical and physical experiments are performed to validate this newly developed nPCT image reconstruction algorithm. Results demonstrate that this proposed method can ensure high-precision nano-resolution phase contrast CT imaging.

In conclusion, a novel model-driven nano-resolution phase contrast CT image reconstruction algorithm is developed for the Lau interferometer based nano-resolution hard X-ray phase contrast imaging.

\section{Funding} 
National Natural Science Foundation of China (12027812), Guangdong Basic and Applied Basic Research Foundation (2021A1515111031), and the Youth Innovation Promotion Association of the Chinese Academy of Sciences (2021362). The experimental data was obtained by Momose's project supported by Exploratory Research for Advanced Technology (ERATO) (JPMJER1403) of Japan Science and Technology Agency.

\section{Acknowledgment} 
The authors would like to thank Dr. Hidekazu Takano for performing the experiments and preparing the raw data. 

\bibliography{sample}

\end{document}